\documentclass[aps,pre,twocolumn,flosuperscriptaddress]{revtex4}
\usepackage{graphicx}
\usepackage{amsmath,amssymb}
\usepackage[utf8]{inputenc}   
\usepackage{color}

\newcommand{\rd}{r_{\scriptscriptstyle\rm D}}
\newcommand{\rc}{r_{\scriptscriptstyle\rm C}}
\newcommand{\fc}{f_{\scriptscriptstyle\rm C}}
\newcommand{\rp}{r_{\scriptscriptstyle\rm P}}
\newcommand{\ri}{r_{\scriptscriptstyle\rm int}}
\newcommand{\rhoc}{\rho_{\scriptscriptstyle\rm C}}

\begin{document}  
\title{Percolation and cooperation with mobile agents: geometric and strategy clusters}
\author{Mendeli H. Vainstein}
\email{vainstein@if.ufrgs.br}
\affiliation{Instituto de F{\'\i}sica, Universidade Federal do Rio
  Grande do Sul CP 15051, 91501-970 Porto Alegre RS, Brazil}
\author{Carolina Brito}
\email{carolina.brito@ufrgs.br}
\affiliation{Instituto de F{\'\i}sica, Universidade Federal do Rio
  Grande do Sul CP 15051, 91501-970 Porto Alegre RS, Brazil}
\author{Jeferson J. Arenzon}
\email{arenzon@if.ufrgs.br}
\affiliation{Instituto de F{\'\i}sica, Universidade Federal do Rio
  Grande do Sul CP 15051, 91501-970 Porto Alegre RS, Brazil}
\date{\today}

\begin{abstract}
We study the conditions for persistent cooperation in an off-lattice model of mobile agents playing the Prisoner's Dilemma game with pure, unconditional strategies. Each agent has an exclusion radius $\rp$ 
 that accounts for the population viscosity, and an interaction radius $\ri$ that defines the instantaneous contact network for the game dynamics. We show that, differently from the $\rp=0$ case, the model with finite sized agents 
presents a coexistence phase with both cooperators and defectors, besides 
the two absorbing phases in which either cooperators or defectors dominate. We provide, in addition, a geometric interpretation of the transitions  between phases.
In analogy with lattice models, the geometric percolation of the contact network (i.e., irrespective of the strategy) enhances cooperation. More importantly, we show that the percolation of defectors is an essential condition for their survival.  Differently from compact clusters of cooperators, isolated groups of defectors 
will eventually become extinct if not percolating, independently of their size.
\end{abstract}

\maketitle

\section{Introduction}

Network reciprocity~\cite{Axelrod84,NoMa92} is a general mechanism 
 responsible for the development of spatial correlations within a viscous population, opening the possibility
of persistent cooperation. Several specific models have been 
proposed showing how these correlations are related to stable groups 
of cooperating individuals, whose bulk benefits of self-defense and mutual 
support outcompete the surface exploitation by
defectors~\cite{DoHa05,Nowak06,SzFa07,RoCuSa09a,PeSz10}. Although actual
experiments have been performed~\cite{TrSeSoKrMi10,RaArCh11,GrFeRuCuSaMo12,GrRoSeMiTr12},
most of our knowledge comes from these simple models. 
In particular, a prevailing characteristic in real systems and an important
ingredient for cooperation is 
the heterogeneous contact in 
systems whose interactions are given by complex~\cite{SaPa05,SaPaLe06} or diluted 
networks~\cite{VaAr01}. 
When we consider  the Prisoner's
Dilemma (PD) dynamics~\cite{Axelrod84} on a diluted lattice that, albeit heterogeneous, has only short range interactions, intermediate densities present an enhancement of cooperation~\cite{VaAr01,LiLiTiSh10,WaSzPe12}
and, in the presence of a small amount of noise, the optimal dilution is
closely related  to the (random site) percolation threshold for that lattice~\cite{WaSzPe12}. 

Whatever the level of heterogeneity, the contact network topology may evolve in time. Although
several rewiring mechanisms can be devised (see Ref.~\cite{PeSz10} and references therein), this may also be accomplished when the high viscosity
restriction is relaxed and the agents become mobile. Mobility patterns on 
different scales of human activity, and their far fetched consequences, have been studied in recent decades. 
For example, airplane displacement and its connection with disease  spread~\cite{CoBaBaVe06}, on a global level, 
can be contrasted with the more local dynamics of pedestrians, crowds or traffic~\cite{HeFaVi00,BoKu14}. Of 
particular interest is how the observed patterns can affect the outcome of the competition between 
agents and, in turn, be influenced by it as well.  
Within the evolutionary game theory framework, after several sparse, early attempts to include 
mobility~\cite{DuWi91,EnLe93,FeMi96,MaLiLiSp99,HaTa05,Koella00,Aktipis04,LeFeDi05}, 
it was only recently that
the interest in the combined effects of mobility and cooperation in the
PD game had a significant increase.
Some level of information processing capability is required,  for example, when the movement is strategy dependent~\cite{JiWaLaWa10,ChLiDaZhYa10} or driven by
payoff~\cite{YaWuWa10,ChDaLiZhZhYa11,LiYaSh11,LiYaWuWa11},
success~\cite{HeYu09,Yu11,LiChZhTaWa12,BuToAn13} or the neighborhood composition~\cite{ChGaCaXu11a,ZhWaDuCa11,ZhZhWePeXiWa12,CoWuQiWa12,ChDaLiQiZhYa13,IcSaSaWi13,ZhZhXi14,AnToBu14}.
However, the simplest scenario is when mobility is 
diffusive~\cite{VaSiAr07,DrSzSz09,SiFoVaAr09,MeBuFoFrGoLaMo09,ChLiDaZhYa10,SuKi11,SmSc12,YaWa11,GeCrFr13}. Indeed, as hypothesized in Ref.~\cite{WeWaLiNeSiWeLe11}, random mobility may have evolved prior to contingent mobility, allowing bacteria to move away from each other while exploring new resources. 
Our previous results on a lattice~\cite{VaSiAr07,SiFoVaAr09,VaAr14} show that even in
the framework of random, noncontingent mobility of unconditional agents,
diffusion is favorable to cooperation, under rather broad conditions, if velocities are not too 
high. Analogous conclusions, attesting the robustness of the results, were also found in off-lattice models~\cite{MeBuFoFrGoLaMo09,AmFo11,AnToBu14}.

When diffusion occurs on a lattice and the one agent per site constraint holds, 
this area exclusion couples the diffusivity of the agents with the free area.
This dependence on density, on the other hand, is not immediately present in off-lattice 
systems with point-like particles~\cite{MeBuFoFrGoLaMo09,AmFo11,LiYaWuWa11,ChDaLiQiZhYa13,AnToBu14}. 
Moreover, while on a lattice
the number of simultaneous interactions is limited by its coordination number,
there is no such restriction on the number of point-like particles within
the range of interaction in off-lattice systems (unless it is explicitly included
as in Ref.~\cite{AnToBu14}). A relevant question concerns the
universal effects of such geometrical hindrance on the emergence and persistence of
cooperation. For example, letting 
the average body size be a coevolving trait, there may
be some evolutive pressure for not too small cooperators because, assuming random
diffusion, a group of small individuals will more easily evaporate from the cluster 
surface. They should not be too large either and, consequently, not be able to evade defectors  and 
avoid exploitation.  Analogously, for defectors, they should neither be too
small in order to stay closer to their prey for longer periods, nor  too large so new,
more promising regions will not be explored. Therefore, one intuitively expects that 
intermediate, optimal sizes may be beneficial to cooperation and thus be selected for. 
Another possible interpretation for an exclusion zone around each agent is its  protected region, and  the resources within. Whatever the interpretation, it is important
to better understand the relevance of area exclusion in these games. As a first
approximation, we consider an effective radius of exclusion, modeled as a hard disk.

Here we study, by explicitly taking into account the excluded area of the agents,
the interplay between geometry, density and mobility on the capability of 
a simple model to sustain cooperation and the question of whether the transitions in this 
class of model have a geometric interpretation. Although the connection between the
threshold of geometric percolation, that is independent of the game dynamics, and   
cooperation has already been reported in Refs.~\cite{WaSzPe12,WaSzPe12b,YaRoWa14}, we also explore the geometry of clusters of cooperators and defectors, 
and the connection between their critical properties and the transition between regions with and without cooperation, thus providing a geometric interpretation of these transitions.

\section{The Model}

\begin{figure}[ht]
\includegraphics[width=7cm]{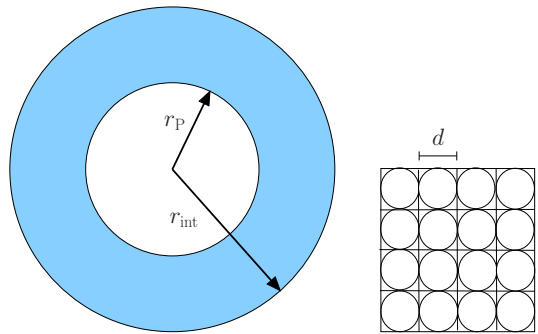}
\caption{(Color online) Geometric parameters of the model, the hard core radius $\rp$ and the interaction
radius $\ri$. While the former sets an exclusion area around each agent, the latter defines
its contact network. The characteristic length $d$, as indicated on the right, is defined by dividing the available area, $L^2$, by the area of the square box around each of the $N$ particles 
of diameter $d$, that is, $L^2/d^2=N$.}
\label{fig.desenho}
\end{figure}

We study an off-lattice model~\cite{NoBoMa94a,MeBuFoFrGoLaMo09,AnToBu14} in which
the $N$ agents living in a square of side $L$ (with periodic boundary conditions) are characterized by an unconditional strategy (cooperate, C, or defect, D) and two independent geometric parameters: an interaction radius $\ri$ and a hard disk 
radius $\rp$ to account for excluded area. The radius $\ri$ determines the neighborhood of each agent
and, as a consequence, its instantaneous contact network. 
The area fraction occupied by the hard disk particles is $\phi=N \pi \rp^2/L^2$.
We use  $d\equiv L/\sqrt{N}$  as our length scale.  These geometric parameters are illustrated in Fig.~\ref{fig.desenho}. The particular case studied by Meloni {\it et al}~\cite{MeBuFoFrGoLaMo09} 
is recovered in the limit of point-like particles, $\rp=0$.

Initially, $N$ individuals with probability 1/2 of being either C or D are randomly placed  in such a way that there is no overlap between any
two individuals $i$ and $j$, i.e., their center-to-center distance $r_{ij}$ satisfies $r_{ij} \geq 2\rp$. Moreover, they
are allowed to randomly diffuse while playing the PD game with their neighbors.  Two agents are considered neighbors if $r_{ij} < \ri$.
A time step is defined as a sequence of $N$ attempts of diffusion and a complete, synchronous round of the PD in
which each of the $N$ agents plays with all its neighbors. During the diffusive part, the
position $(x_i, y_i)$ of the center of particle $i$ at time $t$ is updated if there is no overlap 
between particles in the final position:
\begin{align*}
x_i(t+1) &= x_i(t) + \mu\ri \cos\theta_i(t) \\
y_i(t+1) &= y_i(t) + \mu\ri \sin\theta_i(t).
\end{align*}
Each step has a constant size, $\mu\ri$, and a random orientation $\theta_i(t)$ 
drawn from a uniform distribution in the interval $[-\pi,\pi]$. When $\mu=0$
there is no mobility and we consider here that $\mu$ is small enough so that jumps over
other agents do not occur.   
Under mutual cooperation (defection), both receive payoff $R$ ($P$) as a reward (punishment); 
if one cooperates and the other defects, then the latter receives $T$ (temptation) and the former, $S$. 
To characterize the PD, the following inequalities should hold $T>R>P\geq S$ and $2R>T+S$. In particular, 
we use $R=1$, $P=S=0$ 
and $T>1$, a common parametrization known as the weak form of the PD game.  The evolution follows the finite 
population analog of the replicator dynamics~\cite{SaPaLe06}. Each individual $i$, after accumulating the payoff 
from all combats, randomly chooses a neighbor $j$ with whom to compare their respective payoffs $P_i$ and $P_j$. 
If $P_i \geq P_j$, then $i$ maintains its strategy. On the other hand, if  $P_j>P_i$, $i$ will adopt the strategy 
of $j$ with probability proportional to the payoff difference
\begin{equation}
\Pi_{ij}=\frac{P_j-P_i}{\text{max}\{k_j,k_i\}T},
\end{equation}
where $k_i$ and $k_j$ are the number of neighbors of $i$ and $j$, respectively.  Under this update rule, the 
total number of individuals is kept constant.  

Most of our results are for  $N=32^2$, $T=1.1$, $\mu=0.01$ and $L=1$. We then check the robustness of the model by testing finite size effects with up to $N=128^2$ particles, as well as the dependence on $T$ and $\mu$. Averages are taken over 100 or more samples.

\section{Cooperation and percolation}

\begin{figure}[htb]
\includegraphics[width=7cm]{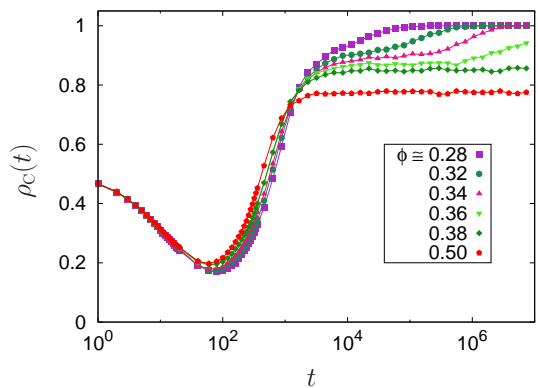}
\caption{(Color online) Average fraction of cooperators as a function of time (in Monte Carlo steps).  
$N=32^2$, $T=1.1$, $\ri= \sqrt{3.5}d$, 
and $\mu=0.01$ for various area fractions $\phi$. Below $\phi\simeq 0.37$, cooperators
eventually invade the whole system. As $\phi$ approaches this threshold, the time spent close to the critical plateau at $\rhoc\simeq 0.85$ also increases.}
\label{fig.temporal_evolution}
\end{figure}

Two macroscopic asymptotic quantities, once averaged, are used to characterize the system: 
the fraction of cooperators $\rhoc$ (those, among the $N$ agents, that cooperate) and 
the fraction $\fc\leq\rhoc$ of initial conditions whose evolution ends in the absorbing state $\rhoc=1$.
 Their difference, $\rhoc-\fc$, is a measure of the coexistence of both strategies.
Four regimes are present in the time evolution, as seen by the behavior of $\rhoc(t)$ in 
Fig.~\ref{fig.temporal_evolution}. As is often the case for this class of model, there is an 
initial drop in the fraction of cooperators from $\rhoc(0)=1/2$, 
since small 
cooperator clusters are easily preyed on in the beginning of the simulation. As $\rhoc(t)$ approaches its minimum value at $t\sim 10^2$, 
fluctuations may lead to extinctions in finite size systems. Away from the minimum, the 
surviving clusters of cooperators resume growth.
These two initial regimes are quite independent
of the occupied area fraction, as indicated in Fig.~\ref{fig.temporal_evolution} 
by the close proximity of all curves up to $t\sim 10^3$. In the third regime, $\rhoc(t)$ attains 
a plateau where it stays, indicating persistent coexistence of both strategies, if $\phi$
is large enough (in the case of Fig.~\ref{fig.temporal_evolution}, the threshold is at $\phi\simeq 0.37$). 
When $\phi$ is below the threshold, after the stasis period on the plateau, the system enters the 
last regime in which cooperators take over the system, that is, $\rhoc(\infty)=1$. The time
to reach this asymptotic state seems to diverge as the threshold area fraction
is approached from below. 

\begin{figure}[htb]
\includegraphics[width=8cm, height=5.5cm]{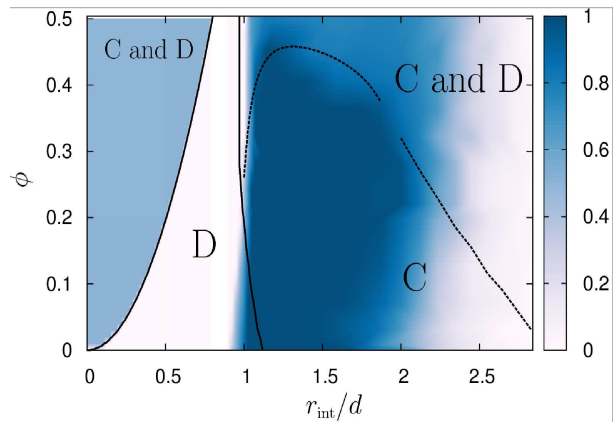}
\caption{(Color online) Phase diagram in the plane $\ri/d$ and $\phi$ for  $N=32^2$, $\mu=0.01$ and $T=1.1$. The
color code is the average fraction of cooperators, $\rhoc$. 
Besides the phases corresponding to the absorbing states, `C' ($\rhoc=\fc=1$) and `D' ($\rhoc=\fc=0$), 
there are two coexistence phases, `C and D'.}
\label{fig.diagram}
\end{figure}

\begin{figure}[htb]
\includegraphics[width=8cm]{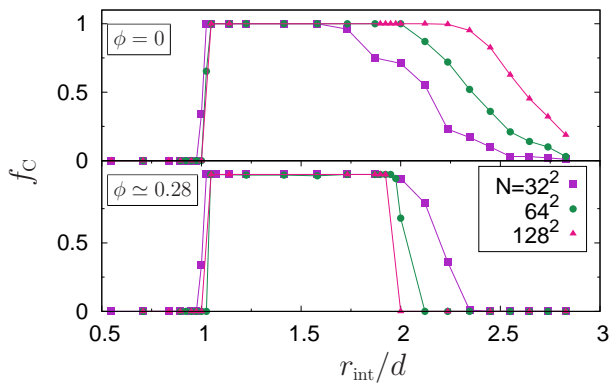}
\caption{(Color online) Fraction of initial states that go to the $\rhoc=1$ absorbing state as a function of $\ri/d$ for several system sizes and two area fractions $\phi=0$ and 0.28. For large values of $\ri/d$, as $\fc$ increases for larger system sizes, the existence of the $\rhoc=0$ state for point particles ($\phi=0$) is due to finite size fluctuations.  For $\phi\simeq 0.28$ the behavior is the opposite and the transition becomes sharper when the system size increases.}
\label{fig.finite_size}
\end{figure}

Fig.~\ref{fig.diagram} summarizes our most important results, showing,  
in the stationary regime, the average fraction of cooperators $\rhoc$ as a function 
of both $\ri/d$ and $\phi$. The lines are independent measures of percolative properties that will be explained
below.  Several regimes may be identified: two absorbing phases in which all agents eventually become either 
defectors ($\rhoc=\fc=0$, labeled `D') or cooperators ($\rhoc=\fc=1$, `C') and  two coexistence ones  
($0<\rhoc<1$ and $\fc=0$, `C and D'). Finite systems may also present a bi-stable phase, 
in which all initial conditions lead to one of the absorbing states.
In this case, although the average cooperativeness still obeys $0<\rhoc<1$, it differs from the coexistence state since $\fc=\rhoc$. However, by increasing the system size, the probability of becoming dominated by defectors goes to zero inside the 'C' region in Fig.~\ref{fig.diagram} and, taking this into account  (see Fig.~\ref{fig.finite_size}), we already properly labeled it. 
We now concentrate on the results for a finite system with $N=32^2$ particles, and discuss the  finite size effects at the end of this section.

There are two limits of the diagram that have trivial results. For very large values of $\ri$, very distant particles interact, increasing the number of contacts and
decreasing the effects of spatial correlation. Thus, we recover the mean 
field result in which all agents become defectors. This trivial region of the phase diagram 
was not explored. In the other limit, when $\ri<2\rp$, the hard core prevents any interaction between the agents and the fraction of 
cooperators remains equal to the initial one, $\rhoc=1/2$ and $\fc=0$. This is the trivial coexistence region, labeled `C and D' on the left of Fig.~\ref{fig.diagram}, above the line $\phi=(\pi/4)(\ri/d)^2$, corresponding to $2\rp=\ri$. Immediately to the right of this line, interaction, albeit weak, is possible and clusters are formed. However, they are small and do not favor cooperation, therefore $\rhoc = 0$, and this all defector phase is labeled `D'.
As we discuss below, the transitions between the other phases have geometric origins and are closely related to the percolating properties of the contact networks.   

\paragraph*{Geometric percolation.}

For the area fractions considered here, non-trivial cooperation first appears around 
$\ri\approx d$. For small $\phi$, the transition between phases D to C corresponds to a change in stability 
of the absorbing state, from the defector to the cooperator dominated phase, 
while for larger $\phi$, roughly $\phi>0.25$, the emergent phase is one in which cooperators and defectors coexist, 'C and D'. 
This onset of cooperation is strongly correlated with the appearance of a 
{\it geometric} percolating cluster, indicated by the steep line in Fig.~\ref{fig.diagram} at the point where the probability of finding a percolating cluster is $50\%$. This is a purely geometric problem of disks with both an inner hard core in addition to a soft, penetrable region, and thus independent of  the game dynamics. 
In other words, the network of contacts of the percolating cluster spans the whole length of the  system.
For $\phi=0$, our result is consistent both with the percolation threshold obtained numerically in Refs.~\cite{QuToZi00,MeMo12},
and the exact bounds of Ref.~\cite{BaBoWa05}. 
For finite $\phi$, the threshold is slightly smaller
than for $\phi=0$, since as $\rp$ increases, the overlap between the disks 
decreases and percolation is attained with a smaller $\ri$. 

\paragraph*{Percolating clusters of defectors.}

\begin{figure}[ht]
\includegraphics[width=8cm]{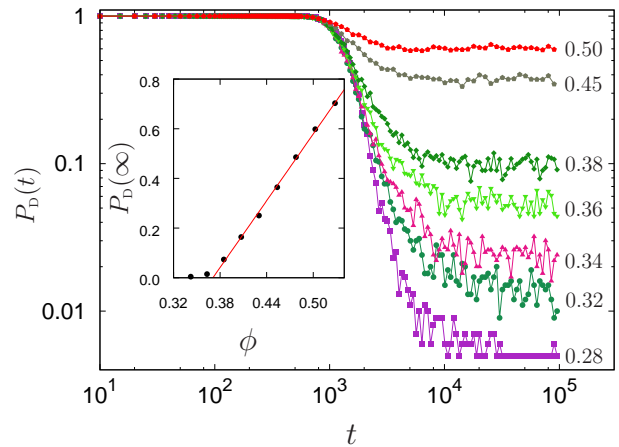}
\caption{(Color online) Probability of percolation of D clusters as a function of time, $P_{\scriptscriptstyle\rm D}(t)$,
for several area fractions $\phi$ and the same parameters of Fig.~\ref{fig.temporal_evolution}. Inset: Asymptotic limit of  $P_{\scriptscriptstyle\rm D}(t)$  as a function of $\phi$. The linear fit indicates that the percolation  probability goes to zero at $\phi\simeq 0.37$. The small deviation seen close to this point is due to the long time of convergence.}
\label{fig.perc}
\end{figure}

Besides the clusters of particles irrespective of their strategies, we also consider the geometry of clusters composed only by defectors (that are, in turn, intimately connected with the geometry of cooperator  clusters), 
which depends on the particular strategy evolving dynamics. Fig.~\ref{fig.perc} shows
the probability of percolation of D clusters as a function of time, $P_{\scriptscriptstyle\rm D}(t)$, for  $\ri=\sqrt{3.5}d$. Notice that in this region there always is a percolating geometric cluster. Initially, as the fraction of cooperators decreases towards the minimum, there is a sea of defectors that obviously percolates and all curves overlap at $P_{\scriptscriptstyle\rm D}=1$. It is only when the curves of $\rhoc(t)$, for different values of $\phi$, start to separate, around $t\sim 10^3$, that $P_{\scriptscriptstyle\rm D}(t)$ starts decreasing. The asymptotic probability of there being a percolating cluster of defectors attains a limiting value, as shown in the inset of Fig.~\ref{fig.perc}, from which the threshold can be obtained. For the particular value of $\ri$ shown in this figure, when the area fraction is below the threshold at (roughly) 0.37, $P_{\scriptscriptstyle\rm D}(\infty)=0$ and, as can be seen in Fig.~\ref{fig.temporal_evolution}, $\rhoc(\infty)=1$.
 The dotted transition line in the phase diagram (Fig.~\ref{fig.diagram}) is obtained in the same manner: it is the  asymptotic value of $\phi$ at which  $P_{\scriptscriptstyle\rm D}(\infty)=0$ as a function of  $\ri$. In the C region, finite size fluctuations sometimes lead to the all-D state $(\ri/d \gtrapprox 2)$, but these configurations are not taken into account for the calculation of the asymptotic value of $\phi$. 
This transition line suggests an important ingredient to understand the coexistence between cooperators and defectors: 
only under the presence of a percolating sea of defectors is that a stable 
coexistence between cooperators and defectors is possible. In other words, differently from compact clusters of cooperators, isolated groups of defectors either grow and percolate or eventually become extinct. 

\paragraph*{Finite size effects.}

We now discuss how the system size can affect the phase diagram, Fig.~\ref{fig.diagram}.
For finite sized systems, a small region inside the 'C' phase has a bi-stable
equilibrium, in which all initial conditions lead to an absorbing state, either $\rhoc=0$ or 1.  The all-D state is due to the fact that the population of cooperators becomes quite small during the initial drop in the first generations and, therefore,  sensitive to fluctuations which occasionally cause extinctions.
 Fig.~\ref{fig.finite_size} shows $\fc$ for two different values of $\phi$ and several system sizes.
 For point particles $(\phi= 0)$, the absorbing all-C state region grows as the system size increases. On the other hand, for $(\phi \approx 0.28)$, the all-C region shrinks. In both cases, however, the 'C' phase width converges to a finite value and the bi-stable region decreases as the system size increases. Furthermore, the size of the 'C' region in the limit of large systems is consistent with the transition line obtained from the percolating defector cluster analysis for $N=32^2$, which 
explains why this phase is not homogeneously painted in
Fig.~\ref{fig.diagram}. Notice that when the system is bi-stable, the average fraction of cooperators is not a good measure since it represents neither one of the final states~\cite{KuRi12}. 
On the other hand, in the coexistence state, both strategies are present in the asymptotic state and, while $0<\rhoc<1$, $\fc=0$. For the $\rp=0$ case studied by Meloni {\it et al}~\cite{MeBuFoFrGoLaMo09}, the $\phi=0$ line in the phase diagram, there is no coexistence phase and the system  eventually enters an absorbing state. 

\paragraph*{Robustness against mobility and temptation.}

\begin{figure}[htb]
\includegraphics[width=8cm]{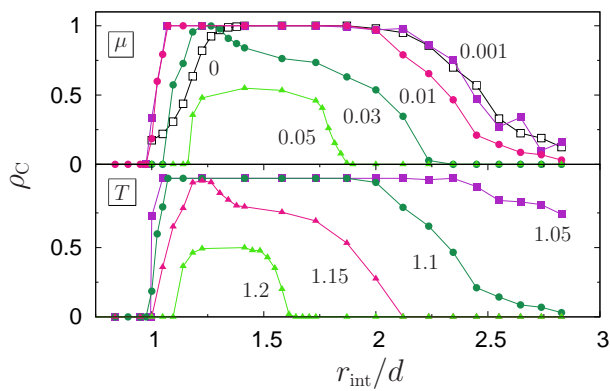}
\caption{(Color online) Average asymptotic fraction of cooperators as a function of $\ri/d$ for  
$\phi\simeq 0.28$ and several values of $\mu$ (top), including $\mu=0$ (empty symbols), 
and $T$ (bottom). In both cases, $N=32^2$; for the top panel, $T=1.1$ while for the bottom one, $\mu=0.01$.}
\label{fig.mu-T}
\end{figure}

We finally consider the robustness of our results when the mobility $\mu$ and the temptation $T$  are varied,
Fig.~\ref{fig.mu-T}. The top panel shows several values of $\mu$, with the temptation fixed at $T=1.1$.  
Whatever the velocity, no cooperation is possible for $\ri\leq d$ due to the absence of a percolating geometric 
cluster. Notwithstanding, for small mobilities, existence of cooperators is possible in a wide range of $\ri$.  When comparing the low mobility case with the one with immobile agents $(\mu=0)$, it can be seen that there is an improvement only for low values of the $\ri/d$ ratio. For high values of this ratio, the curves overlap, which is expected, since the agents have a large neighborhood that is minimally perturbed by the small random movements. On the other hand, for low values of the ratio,   the contact range is small and is more affected by the diffusion.  
Random movements lead to the evaporation of cooperative clusters and, as the velocity increases, 
cooperation levels decrease until a threshold above which it is no longer possible. This effect of cooperation enhancement driven by a low mobility is in accordance with previous simulations on a lattice~\cite{VaSiAr07,SiFoVaAr09,MeBuFoFrGoLaMo09,VaAr14}. 
 The bottom panel of Fig.~\ref{fig.mu-T} shows the effect of the parameter $T$. As expected, as the temptation to defect increases, the fraction of cooperators decreases.

\section{Discussion and conclusion}

We presented numerical results for an off-lattice model of mobile agents
playing the PD game while randomly moving across a closed region. This model
combines ingredients found in two distinct models for such systems: the
excluded area found in lattice simulations and an available continuous 
space. We arrive at two important results: first, if the agents are not
point-like, a non-trivial coexistence phase with cooperators and defectors 
becomes possible (besides the trivial one when the exclusion range
is larger than the interacting radius); second, it is possible to geometrically
interpret, in terms of percolation, the observed transitions. An important role is played not
only  by geometric percolation, irrespective of the game dynamics, but also by the percolative properties of defector clusters, whose threshold depends on the details of the game. 
 
When the agents interaction is prevented by the hard core, trivial coexistence
between cooperators and defectors is possible. Beyond that region of the phase
diagram, no cooperation is possible in the absence of a percolating geometric 
cluster. Random mobility provides an
evaporating mechanism for groups of cooperators what is detrimental to cooperation
as isolated cooperators are easily preyed on. However, in the presence of a
percolating clusters, it is easier for a detaching cooperator to be in contact
with another cluster and be protected. When hard cores are included, movements
are hindered and the agents spend some time rattling around the same
region, while large displacements become less probable, what increases the correlation
among agents and benefits cooperation even further. This localization is
probably the mechanism responsible for the coexistence between cooperators and
defectors that is not present for $\phi=0$. Interestingly, the presence of defectors 
is only possible if they form a percolating cluster and no finite cluster of defectors 
is stable: they either grow, merge with others and span the whole lattice or the isolated cluster becomes extinct. This is our main result: finite size cooperators and defectors, 
whose hard core is an effective, averaged interaction restraining their movements,
are able to coexist over a broad region of the phase diagram only if  defectors are organized in a interconnected cluster, a percolating sea of defectors. A similar effect was found for the public goods game
played on a lattice with empty sites and no mobility~\cite{WaSzPe12}.
It would be interesting to investigate whether this  condition
for coexistence between cooperators and defectors 
also occurs in lattice models, where excluded area is inherent to the formulation of the problem.

In this manuscript we focused on the particular homogeneous case of equal sizes and equal velocities
for both cooperators and defectors.  Following Refs.~\cite{VaAr14,AnToBu14},
it is essential to explore the whole $(S,T)$ parameter space and the dependence on the chosen dynamic
rule in order to check the robustness of cooperation. Furthermore, several extensions are possible. For example,
velocities may not be constant~\cite{ChLiDaZhYa10} and depend on the neighborhood~\cite{AnToBu14} or strategy.  
The hard core radius may also correlate with strategy, $\rc$ and $\rd$ for cooperators
and defectors, respectively. In particular, if individuals coevolve with mutations: is there an optimal equilibrium radii ratio to which the system converges to or, instead, a permanent arms race? What are the effects of having size dispersion? If velocity and size coevolve along with 
strategies, defectors may become small and fast while cooperators  become large and slow.
Finally,
what happens if there is a fraction of
fundamentalists (both cooperators and defectors or just defectors) whose strategies or positions never change? 
In all these cases, it is important to study also the geometric properties of the interfaces between
cooperator and defector clusters since these are the places where all strategy flips occur.
From a more physical perspective, it would be interesting
to find out, for each transition line, which is the dynamical universality class that it belongs to~\cite{YaRoWa14}.
These and other relevant questions are being considered.

\begin{acknowledgments}
Research partially supported by the Brazilian agencies CNPq,
CAPES and Fapergs.  JJA also thanks the INCT-Sistemas Complexos (CNPq) for
partial support. We thank the supercomputing laboratory at Universidade Federal da Integração Latino-Americana (LCAD/Unila), where the simulations were run, for computer time.
\end{acknowledgments}


%

\end{document}